\documentclass[a4paper,12pt]{article}%
\usepackage{amsmath}%
\usepackage{txfonts}%
\usepackage{amsfonts}%
\usepackage{amssymb}%
\usepackage{graphicx}
\usepackage{wrapfig}
\usepackage{color}
\usepackage{setspace}
\usepackage{esint}
\usepackage[rightcaption]{sidecap}
\usepackage{wrapfig}
\usepackage{lipsum}
\usepackage{caption}
\usepackage{longtable}
\usepackage[T2A,T1]{fontenc}
\usepackage[utf8]{inputenc}
\usepackage[russian,english]{babel}

\usepackage{etoolbox}
\makeatletter
\patchcmd{\@maketitle}{\begin{center}}{\begin{flushleft}}{}{}
\patchcmd{\@maketitle}{\begin{tabular}[t]{c}}{\begin{tabular}[t]{@{}l}}{}{}
\patchcmd{\@maketitle}{\end{center}}{\end{flushleft}}{}{}
\makeatother

\numberwithin{equation}{section}

\newcommand{\be}{\begin{equation}}
\newcommand{\ee}{\end{equation}}

\usepackage{anysize}
\marginsize{2cm}{2cm}{2cm}{2cm}
\usepackage{mathptmx}
-1in\relax 
\begin{document}

\begin{center}
\LARGE
\textbf{Possible Material Outcomes of Solar Power Revolution}
\normalsize
\vskip1cm
\begin{tabular}{l}
\hskip1cm
\Large Mikhail V. Shubov         \\
\normalsize \\
%University of New Hampshire       &
\hskip1cm University of MA Lowell    \\
%33 Academic Way                   &
\hskip1cm One University Ave,        \\
%Durham, NH 03824                  &
\hskip1cm Lowell, MA 01854           \\
%E-mail: marianna.shubov@gmail.com &
\hskip1cm E-mail: mvs5763@yahoo.com  \\
\end{tabular}
\end{center}

\begin{center}
  \textbf{Abstract}
\end{center}
\begin{quote}
Energy production plays a primary role in industrial capacity and thus material standard of living of any civilization.  The Industrial Revolution was engendered by a vast growth of motive energy production.
Solar Power Revolution has the potential of increasing global energy production by a factor of 160 and engendering a new technological revolution in industry, food production and transportation.
Electric energy can be used for vast expansion of industry.
Electric energy can be used to extract vast amount of hydrocarbon motor fuel and chemicals from unconventional oil resources and coal.  It can also be used to produce liquid hydrocarbon fuel, chemicals, and animal feed from water and carbon dioxide.
Electric energy can be used to connect the World by a network of rapid transit, such as Magnetic Levitation Trains (MagLev).
The Solar Power Revolution will enable Earth to sustain a population of 50 billion at material living standards much higher than corresponding standards in USA 2020.  Global civilization which harvests the maximum possible fraction of solar energy falling on Earth is called Kardashev 1 Civilization.  It is also the stepping stone toward the Final Frontier of Humankind -- Kardashev 2 Civilization, which has colonized the Solar System.
\end{quote}

\section{Introduction}
In every civilization, energy is the most important resource for industry.  Availability of energy is the primary factor determining any civilization's industrial capacity and thus material standard of living \cite{E1}.  Availability of motive energy was one of the main factors enabling Industrial Revolution and complete transformation of material living standards \cite{ERev1,ERev2}.  Many experts agree that our civilization is on the brink of a new Power Revolution propelled by solar energy \cite{SRev1,SRev2}.

The Future remains a mystery.  Nevertheless, forecasting of technological developments based on well-established patterns and available resources is possible \cite{Future1,Future2}.  In this work, we estimate energy resources available for Solar Power Revolution.  We estimate energy production by Global Civilization which has mastered Solar Energy.  We also estimate power consumption by different branches of industry.

In Section 2, we discuss modern energy reserves and consumption.
In Section 2.1, we describe different types of energy and their corresponding values.
In Section 2.2, we present some statistics on the historical growth of energy use.
In Section 2.3, we quantify annual energy use in the USA --  24 $PWh$ thermal and  4 $PWh$ electric.
In Section 2.4, we quantify annual global energy use -- 147 $PWh$ thermal and 26 $PWh$ electric.
In Section 2.5, we present a few energy prices.
In Section 2.6, we describe the conventional resources of fossil fuel and uranium.

In Section 3, we describe promising sources of energy.
Solar energy described in Section 3.1 has a potential to generate 23,000 $PWh$ electric.
Wind energy described in Section 3.2 has a potential to generate at least 700 $PWh$ electric.  Even though the potential of wind power is much lower than that of solar power, it can play a very important role in early stages of Solar Power Revolution.
Nuclear and Thermonuclear energy described in Section 3.3 and 3.4 present a potential which has not been realized.

In Section 4, we describe non-conventional resources of liquid fuel.  These resources can not produce an Energy Revolution.  Extracting these resources is a very energy-expensive process.  Nevertheless, these resources can provide liquid hydrocarbons for applications where they are irreplaceable, such as motor fuel and chemical feedstock.  In Section 4.1, we describe unconventional oil resources.   Canada and Venezuela contain a total of 360 billion tons of heavy oil.  Russia and USA contain oil shale which can yield 1,400 billion tons of oil.  In Sections 4.2 and 4.3, we describe coal liquefaction.  In Subsection 4.4, we describe production of hydrocarbon fuel and animal feed from carbon dioxide, water, and electric energy.

In Section 5, we describe the effect of Solar Power Revolution on material living standards of society.
In Subsection 5.1, we present a case that all people of the World can live at high material standards.
In Subsection 5.1, we describe the material standards of Global Civilization after Solar Power Revolution.
This Civilization can support a population of 50 billion at material living standards considerably exceeding those in USA, 2020.
This Society is called a Kardashev 1 Civilization, since it utilises solar energy striking the Earth to a maximum possible extent.
In Subsection 5.3 we describe the colonization of Solar System and Kardashev 2 Civilization which uses most power produced by the Sun.  This civilization is vastly greater than Kardashev 1 Civilization.  It can support a population of 10 quadrillion at material standards vastly superior to those of Kardashev 1 Civilization.

Section 6 is conclusion.  We reiterate the fact that the Future remains a mystery which can not be foreseen.  Nevertheless, by accessing energetic and material resources available on Earth and within the Solar System, we can have a basic road map of future technological progress.

\section{Energy Consumption and Reserves, 2020}
\subsection{Different Types of Energy}
Prior to discussing the future prospects of Energy Revolution, we describe several types of energy and their values.  One type of energy is the electric energy available from regular power grid.  The price for grid electric energy for households is generally about twice as high as the price of grid electricity for industry.  Both of these prices vary by country.  For instance, the average price of residential electricity in 2017 was 13 cents per $kWh$ in the USA and 35 cents per $kWh$ in Germany \cite[p.53]{WEnrg18}.

Another type of energy is thermal energy.  In USA 2019, power stations bought fuels for the following average prices:
coal for 0.70 cents per $kWh$ thermal
natural gas for 1.0 cents per $kWh$ thermal
distillate fuel oil for 5.3 cents per $kWh$ thermal
premium gasoline for 9.4 cents per $kWh$ thermal
\cite[p.158,165]{USAEnergy}.
Effective costs of propulsive energy of most vehicles are much higher.

Some forms of energy have truly exorbitant costs.  The cost of destructive energy delivered to the target by theater missiles is one example.  An ATACMS missile can precisely deliver a 230 $kg$ warhead to a target 300 $km$ away \cite{atacms}. The average unit cost of ATACMS missile is about \$1,500,000 \cite{atacms}.  ATACMS delivers about 250 $kWh$ of destructive energy to the target at a cost of \$6,000 per $kWh$.  A $kWh$ of energy in the form of antimatter costs \$12,600,000 \cite[p.191]{AntiMatter}.  These exotic forms of energy are important for special applications, but they are not discussed in this work.

\subsection{Historical Growth of Energy Use}
Since the beginning of Industrial Revolution in early 19$^{\text{th}}$ century, energy consumption rapidly grew worldwide.  This is especially true for motive energy and later electrical energy.  In pre-Industrial work, motive energy consumption per capita was about 200 $kWh$ per year, of which 140 $kWh$ came from work animals, 30 $kWh$ came from water wheels, and 30 $kWh$ came from wind power \cite{Shubov1}.
In 2017, US motive energy consumption was 20,400 $kWh$ per capita. Of that energy, 12,600 $kWh$ per capita was electricity \cite[p.69]{WEnrg18} and about 7,800 $kWh$ per capita is gas engine work \cite[p.7-2]{Trnsp18}.
Global motive energy consumption was 4,400 $kWh$ per capita. Of that energy, 3,150 $kWh$ per capita is electricity \cite[p.61]{WEnrg18} and about 1,250 $kWh$ per capita is gas engine work \cite{WEnrg18}.

In some areas, energy consumption growth has been more rapid.
Between 1849 and 1955, the total power of prime movers used in American Industry and transportation grew by a factor of 840 \cite[p. 503]{HStat2}.

Recently, the growth of energy use has slowed down, but it is still appreciable.
Between 1973 and 2017, global primary energy consumption grew from 71 $PWh$ thermal to 162 $PWh$ thermal \cite[p.8]{WEnrg18}.  Electric power generation grew from 6.1 $PWh$ to 25.6 $PWh$ during these years \cite[p.30]{WEnrg18}.  Electric power generation efficiency grew from 32.7\% to 37.0\% \cite{EEff}.

\subsection{Energy Use in the USA}
In 2018, USA consumed 23.8 $PWh$ thermal worth of fossil fuels --
 3.9 $PWh$ thermal worth of coal,
 9.1 $PWh$ thermal worth of natural gas,
10.8 $PWh$ thermal worth of petroleum \cite[p.7]{USAEnergy}.
Overall, 4.0 $PWh$ electricity was generated.  Most electricity was generated by fossil-fuel powered plants -- 1.15 $PWh$ by coal, 1.47 $PWh$ by natural gas, and 0.10 $PWh$ by other fuels.  Additionally, 0.81 $PWh$ was generated by nuclear power, 0.29 $PWh$ by hydroelectric power, 0.064 $PWh$ by solar power, and 0.27 $PWh$ by wind \cite[p.127]{USAEnergy}.
About
3.5 $PWh$ thermal worth of coal and
3.4 $PWh$ thermal worth of natural gas
was used in the process \cite[p.213]{USAEnergy}.

In 2018, about 8.35 $PWh$ thermal was used in transportation -- almost all coming from petroleum \cite[p.35]{USAEnergy}.  About 1.7 $PWh$ thermal worth of fuel was used non-combustively.  Mainly, it was used as a feedstock for chemicals \cite[p.35]{USAEnergy}.

About 6.8 $PWh$ thermal is the energy unaccounted for in the previous paragraphs.  This energy was used mostly for residential and commercial space heating, as well as industrial process heating.

\subsection{Global Energy Use}

In 2017, the World consumed 147 $PWh$ thermal worth of fossil fuels --
 44.0 $PWh$ thermal worth of coal,
 36.0 $PWh$ thermal worth of natural gas,
 51.6 $PWh$ thermal worth of petroleum, and
 15.4 $PWh$ thermal worth of waste and biofuel
  \cite[p.47]{WEnrg18}.

Overall, 25.6 $PWh$ electricity was generated.  Most electricity was generated by fossil-fuel powered plants -- 9.9 $PWh$ by coal, 5.9 $PWh$ by natural gas, and 0.84 $PWh$ by petroleum.  Additionally, 2.6 $PWh$ was generated by nuclear power, 4.2 $PWh$ by hydroelectric power, 0.44 $PWh$ by solar power, 1.13 $PWh$ by wind, and 1.6 $PWh$ by other renewables \cite{WEnrg18}.
About
27 $PWh$ thermal worth of coal,
14 $PWh$ thermal worth of natural gas, and
2.5 $PWh$ thermal worth of petroleum
was used in the process \cite[p.47]{WEnrg18}.

In 2017, about 32.6 $PWh$ thermal was used in transportation -- almost all coming from petroleum \cite[p.47]{WEnrg18}.  About 10.2 $PWh$ thermal worth of fuel was used non-combustively.  Mainly, it was used as a feedstock for chemicals \cite[p.47]{WEnrg18}.

About 61 $PWh$ thermal is the energy unaccounted for in the previous paragraphs.  This energy was used mostly for residential and commercial space heating, as well as industrial process heating.

Global flows of thermal energy have been calculated for 2005.
 8.4 $PWh$ thermal went into chemical feedstocks,
25.4 $PWh$ thermal was used by transportation,
24.5 $PWh$ thermal was used by industry, and
33.3 $PWh$ thermal was used by residential and commercial sectors \cite[p.45]{GEAssessment}.

\subsection{Energy Prices}
Energy prices in the USA are plotted below
\cite[p.15]{USAEnergy}:
\begin{center}
\includegraphics[width=16cm]{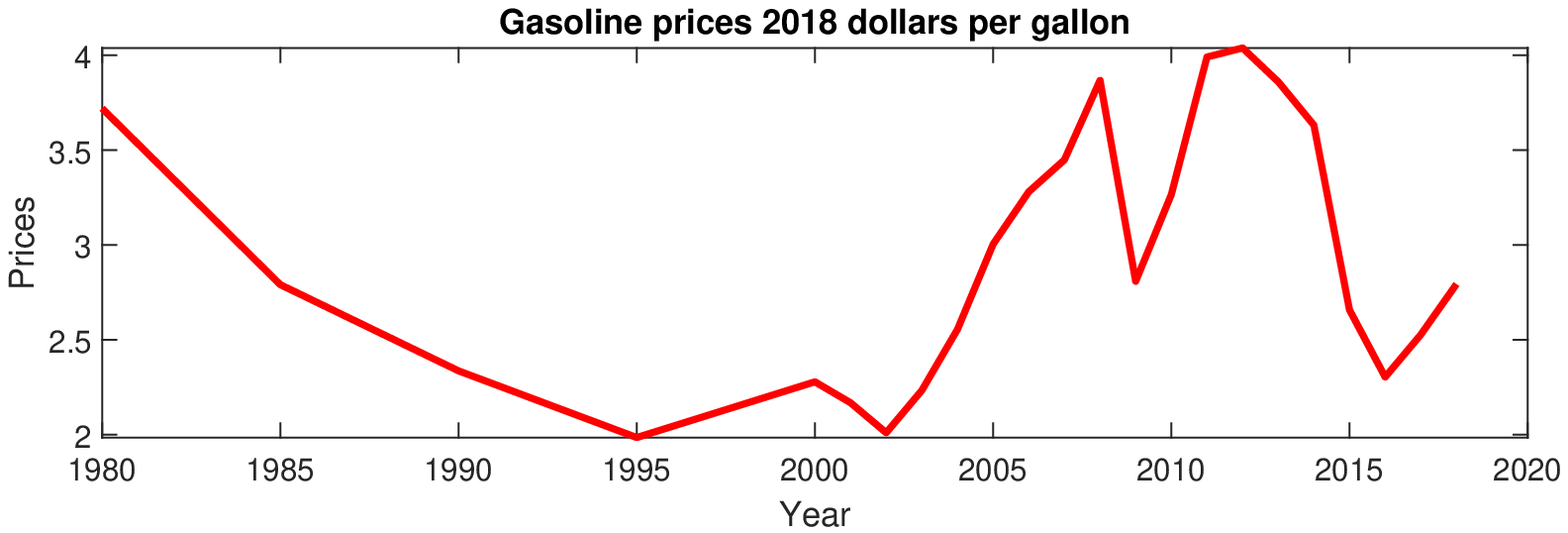}\\
\includegraphics[width=16cm]{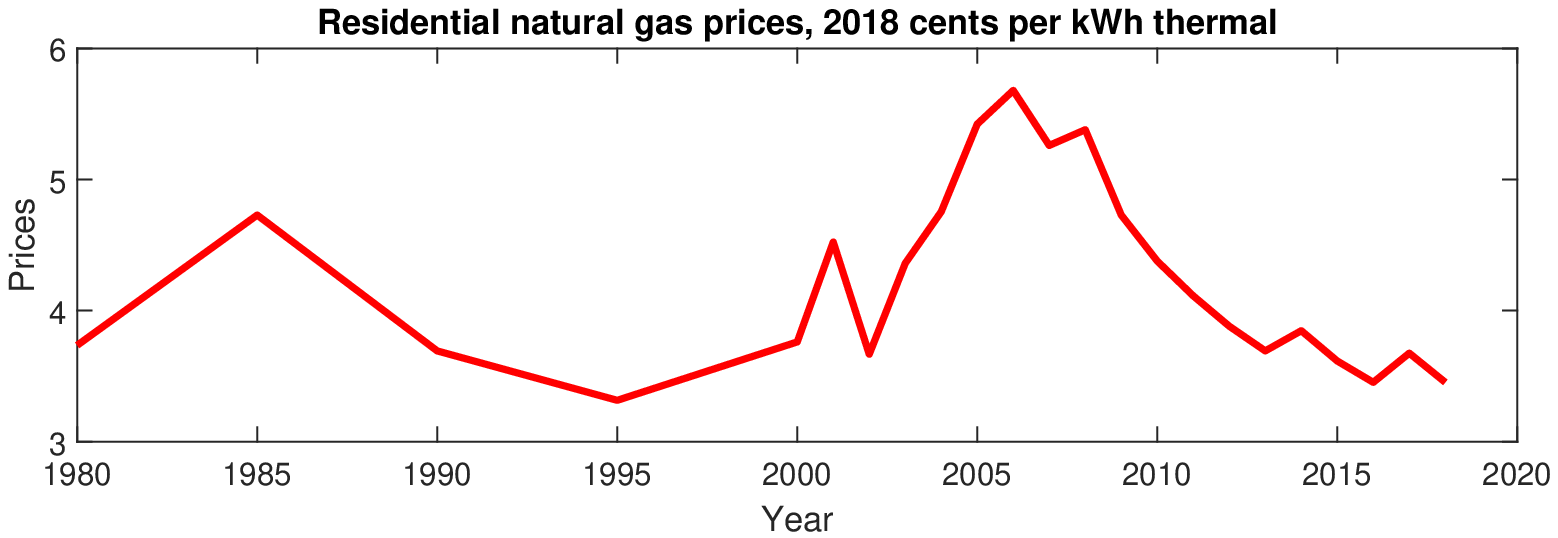}\\
\includegraphics[width=16cm]{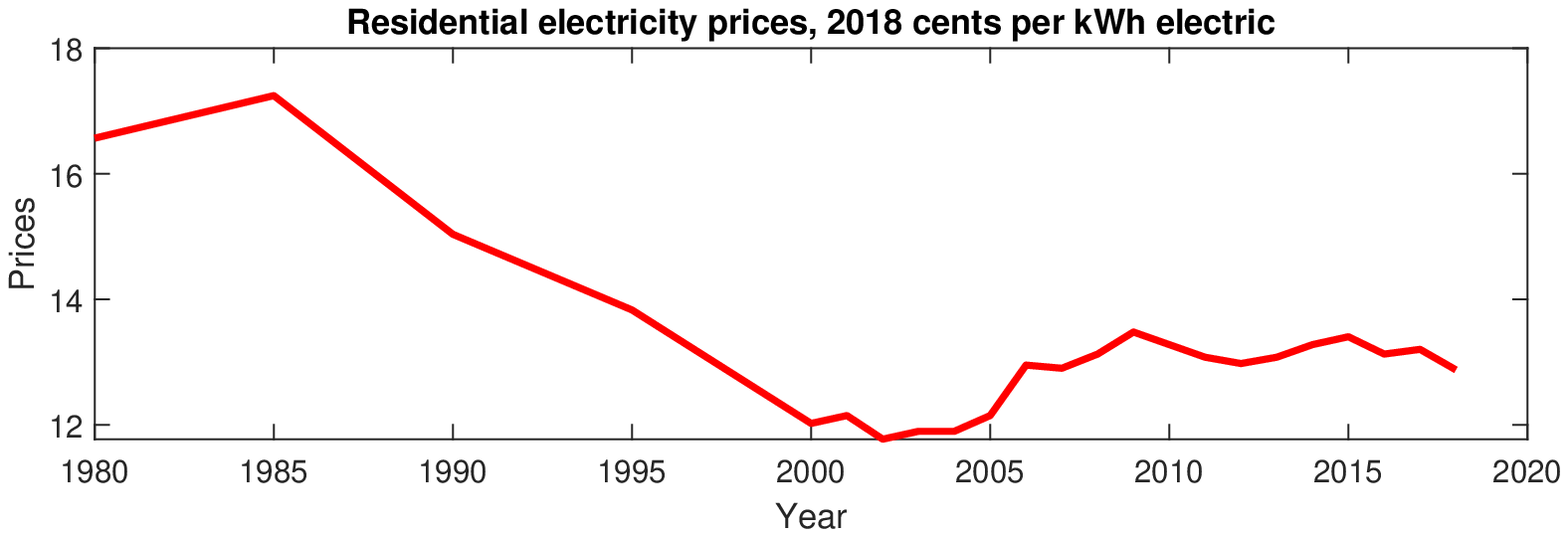}
\captionof{figure}{Energy Prices in the USA \label{2.F01}}
\end{center}

In most places in 2018, gasoline prices are 1.7 to 2.4 higher than in the USA.  Gas prices in different nations are presented in the figure below \cite{GasPrice1}:
\begin{center}
\includegraphics[width=12cm]{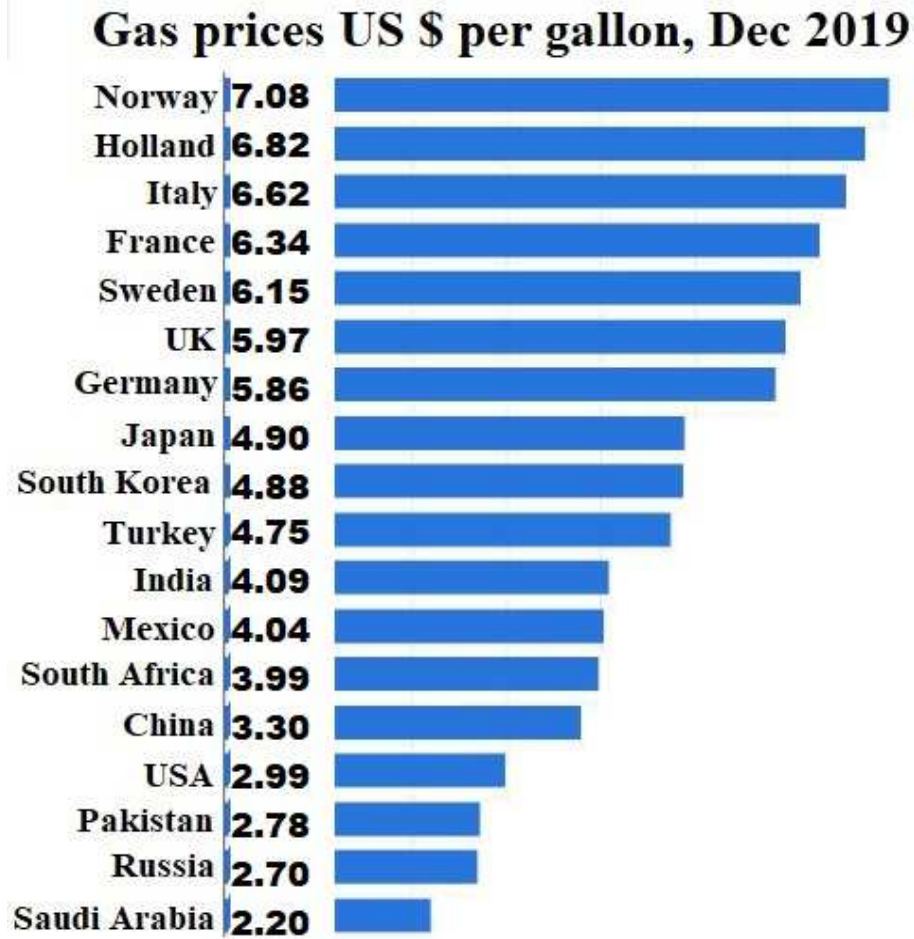}
\captionof{figure}{Gas Prices in December 2019 \label{2.F02}}
\end{center}

Electricity prices are as non-homogeneous as gas prices.  In 2018, in the USA, electricity for industry had an average cost of 7 $c/kWh$.  In most other countries, it was 8 $c/kWh$ to 16 $c/kWh$ \cite[pp.52-53]{WEnrg18}.

\subsection{Conventional Energy Reserves}
Total conventional fuel reserves are 236 billion ton oil, 145 billion tons natural gas, and 1,055 billion tons coal \cite[p.16]{WorldEnergy}.  Reserves of uranium extractable at up to \$260 per $kg$ are 7.8 million tons \cite[p.15]{U2018}.
The energy content of these reserves is the following:
2,500 $PWh$ thermal in oil,
2,070 $PWh$ thermal in natural gas,
6,210 $PWh$ thermal in coal, and
1,060 $PWh$ thermal in uranium.
Energy production is 140 $PWh$ thermal per year \cite[p.17]{WorldEnergy}.

The amount of proven reserves have grown significantly between 1998 and 2018.
Proven oil reserves grew from 156 billion tons to 236 billion tons.
Proven natural gas reserves grew from 96 billion tons to 145 billion tons \cite{WEn19}.

\section{Promising Sources of Energy}
\subsection{Solar Energy}
The technology which has a potential for total transformation of energy production is harvesting of Solar Power.  In order to understand the possible impact of Solar Power Revolution, we must compare the energy produced in Modern World to electric energy which can be produced by Solar Power.
The highest efficiency for a commercial Solar module in early 2020 is 22.6\% \cite{Maxeon}.  We can assume that by the middle of the century, most commercial panels will be at least 24\% efficient.
If all of Earth's deserts are covered with 24\% efficient photovoltaic cells, then the total electricity production would be 7,500 $PWh$ electric per year.

Floating Solar Power is a very promising technology.  It consists of solar power stations floating on water.  Currently, only 0.4\% of all Photovoltaic power is produced by floating solar power stations \cite{FSolar}.  By the end of Solar Power Revolution, Floating Solar Power may become the main energy source.  If 20\% of World Ocean is covered by 24\% efficient photovoltaic cells, then the total electricity production would be 15,500 $PWh$ per year.  This is twice as much as we can obtain from deserts.  All deserts and 20\% of ocean can bring 23,000 $PWh$ per year.

In 2017, worldwide, Solar Power produced about 2.5\% of global electricity.  That year, 0.53 $PWh$ electric was produced by solar power \cite[p.76-77]{PV17}.  In 2018, 0.89 $PWh$ electric was produced by solar power \cite{Solar2020}.  Solar Power production has been growing by an average of 44\% per year since 1992.  It has been growing by an average of 32\% between 2012 and 2017 \cite[p.82]{PV17}.

The cost of installed photovoltaic power has fallen rapidly.  Between 2010 and 2018, the cost of installed solar power for utility-scale stations fell from \$4.63 per $Watt$ to \$1.06 per $Watt$ \cite[p.viii]{PVCost}.  During the same time, the prices of solar modules themselves dropped from \$2.47 per $Watt$ to \$0.47 per $Watt$.  The main breakthroughs came between 2010 and 2013 and in 2016 \cite[p.43]{PVCost}.  By December 2019, most module prices fell to \$0.28 per $Watt$.  Electric energy produced by Solar Power Stations has an average production cost of 5 cents per $kWh$.  Cost decrease has surpassed the 2020 target \cite[p.39]{PVCost}.  Between 2010 and 2018 the average efficiency of the new photovoltaic modules installed in utilities in California grew from 13.8\% to 19.1\% \cite[p.5]{PVCost}.

If the growth rate of 20\% per year can be sustained for 20 years, then Solar Power would produce most electric energy by 2040.  That year, about 40 $PWh$ electric energy should be produced by Solar Power.  The most likely scenario is that after that, Solar Power production will continue to grow.  This would mean the growth of overall power production.  This will likely drive the Second Industrial Revolution.  The growth of Solar Power will continue until it will reach the natural limit of 23,000 $PWh$ described above.

\subsection{Wind Energy}
Like Solar Energy, Wind Energy has a vast potential.  About 1\% of all solar energy striking Earth is turned into wind.  Global wind energy potential is 690 $PWh$ electric per year \cite[p.10,935]{Wind2}.  Other sources estimate it as 870 $PWh$ electric per year \cite[p.25]{Wind3}.  Others provide estimates for offshore potential as 157 to 631 $PWh$ electric per year \cite[p.776]{Wind5}.

Even though the total potential of Wind Energy is much lower than that for Solar Energy, Wind Energy may play an important role during initial stages of the new Energy Revolution.  In 2017, worldwide, Wind Power produced about 6.5\% electric energy \cite[p.77]{PV17}.

Building a power system based on wind energy is a grand undertaking.  Wind turbine generators producing the peak power of 1 $TW$ require the following materials.  An onshore array requires 92 million tons of steel, 6.8 million tons of fiberglass-epoxy composite, 2 million tons of copper and aluminum, and 380 million tons of concrete.
A close offshore array requires 243 million tons of steel, 12 million tons of fiberglass-epoxy composite, 2 million tons of copper and aluminum, and 7.8 million tons of concrete.  A far offshore array would need 550 million tons of steel and the same amount of other material as a close offshore array.
Given the energy requirement for production and transportation of aforementioned materials, it would take 2.2 $PWh$ of electrical energy and energy contained within diesel fuel to build a 1 $TW$ wind array onshore.  It would take 2.4 $PWh$ to build 1 $TW$ wind array close offshore and 3.8 $PWh$ to build 1 $TW$ wind array far offshore \cite[p.53]{Wind8}.  An average 1 $TW$ array would produce 1.9 $PWh$ per year \cite[p.57]{Wind8}.

Under the scenario discussed in \cite[p.58]{Wind8}, 115 $TW$ of wind turbines would be installed onshore and 28 $TW$ offshore.  These turbines would yield 210 $PWh$ electric energy per year.
This project would use 21 billion tons of steel,
44 billion tons of concrete,
1.1 billion tons of fiberglass-epoxy composite,
290 million tons of copper and aluminum.
This immense amount of material can be produced along with energy generated by initial wind turbines.
Global cement production in 2016 was 4.2 billion tons \cite[p.465]{Cement1}.
Global steel production in 2018 was 1.81 billion tons \cite{WSteel}.

Some areas with plentiful wind energy are very remote.  Strongest and most stable winds flow over oceans.  Oceans surrounding Antarctica have the strongest winds \cite[p.40]{Wind8}. Wind turbines in these areas would have to be supported by floating stations.  The energy may be used for floating factories performing energy-intensive tasks.

\subsection{Nuclear Energy}
Nuclear energy held a great promise from 1950s to 1980s, but it did not come true.  The first nuclear power plant in the USA was built by 1957.
By 1970, 20 nuclear power plants operated in the USA, which grew to 71 plants by 1980 and 112 plants by 1990 \cite[p.271]{NUCL01}.  Electricity generation by the nuclear power plants grew even more rapidly.  In 2017, there were 99 nuclear reactors in the USA (less than in 1990), and 449 nuclear reactors in the World \cite[p.70]{U2018}.

In 1957, the sole nuclear power plant in the USA generated 0.2 $TWh$ electric.  By 1970, American nuclear power plants generated 22 $TWh$ electric, which grew to 0.25 $PWH$ electric by 1980, and 0.58 $PWh$ electric by  1990 \cite[p.273]{NUCL01}.  Continued growth of nuclear power production could have started the Energy Revolution.  Nuclear Power Revolution could have started in the 1990s and continued during the first decades of this Century.  Unfortunately, the Nuclear Power Revolution came to an abrupt end before it really started.  Nuclear share of total net generation has not changed much since 1988 \cite[p.273]{NUCL01}.  In 2016, nuclear power plants generated 0.81 $PWh$ electric in the USA and 2.71 $PWh$ electric globally \cite[p.70]{U2018}.

In order to understand the fizzling of Nuclear Power, we must have basic understanding of nuclear reactors.  There are several types of nuclear reactors. The author's paper \cite{MyMasters} was on the subject of Accelerator Breeder Reactors.  Other reactor types relevant to this article are Thermal Reactors and Fast Breeder Reactors, discussed in paragraphs below.

In all nuclear reactors, a chain nuclear fission reaction is sustained.  When a \textbf{fissile nucleus} absorbs a neutron, it is likely to undergo a nuclear fission event.  Examples of fissile nuclei are $^{233}$U, $^{235}$U, and $^{239}$Pu.  A nuclear fission produces several secondary neutrons.  The average number of secondary neutrons produced depends on the energy of the absorbed neutron and the nucleus undergoing fission.  Generally, the average number of secondary neutrons per fission is 2.4 to 2.9.  Some secondary neutrons are absorbed by nonfissile nuclei, while others cause further fission reactions.  In a sustained nuclear fission, the number of neutrons absorbed is almost exactly the same as the number of neutrons produced.  The total neutron flux changes very little over time.

In Thermal Nuclear Reactors, the neutrons are slowed down before they cause a nuclear fission.  Neutrons can be slowed down by multiple collisions with nuclei.  Thermal reactors are by far the most common ones.  In Fast Breeder Reactors, the chain reaction is sustained by fast neutrons.  In Accelerator Breeder Reactors, the nuclear chain reaction is not self-sustaining.  This reaction is sustained by an external source of neutrons.  That source of neutrons consists of a uranium target subject to a stream of superfast protons.  These protons have energy of about 1 $GeV$.  This proton stream is produced by an accelerator.  Whenever a superfast proton strikes a heavy nucleus, it causes the nucleus to disintegrate into many light fragments and neutrons \cite[p.8-13]{MyMasters}.

All reactors consume fissile nuclei such as $^{235}$U, $^{233}$U, and $^{239}$Pu.  Most reactors  also produce fissile nuclei from \textbf{fertile nuclei}.  Examples of fertile nuclei are $^{232}$Th and $^{238}$U.
When $^{232}$Th absorbs a neutron, it becomes $^{233}$Th, which decays to $^{233}$U -- a fissile nucleus.
When $^{238}$U absorbs a neutron, it becomes $^{239}$U, which decays to $^{239}$Pu -- a fissile nucleus.

In Thermal Nuclear Reactors, consumption of fissile nuclei greatly exceeds production of fissile nuclei from fertile nuclei.
In Fast Breeder Reactors (FBR), and more so in Accelerator Breeder Reactors (ABR), production of fissile nuclei from fertile nuclei considerably exceeds consumption of fissile nuclei.
As a result, Thermal Nuclear Reactors must use the resources of fissile nuclei.  Fast Breeder Reactors and more so in Accelerator Breeder Reactors can use the resources of fertile nuclei.  Fertile nuclei are much more common in nature than fissile nuclei.  The only naturally occurring fissile isotope is $^{235}$U, which makes up 0.7\% of all uranium found in nature.  The rest of natural uranium is fertile $^{238}$U \cite[p.6]{MyMasters}.

Fuel reserves for FBRs and ABRs are virtually unlimited.  Fuel reserves for Thermal Nuclear Reactors are very limited.  As we have mentioned in Subsection 2.6, reserves of uranium extractable at up to \$260 per $kg$ are 7.8 million tons \cite[p.15]{U2018}.   Energy content of uranium is 10 times less than that in fossil fuel.  In 2019, Thermal Nuclear Reactors consumed $^{235}$U contained in 67,000 $tons$ of uranium \cite{WorldNuclear}.

No Accelerator Breeder Reactors have been built.  In December 2019, there are 444 nuclear reactors in the World with total power of 395 $GW$ \cite{WorldNuclear}.  There are also 6 Fast Breeder Reactors in the World with total power of 2 $GW$ \cite{WorldNuclear1}.  Fast Breeder Reactors held a promise of providing unlimited energy supply \cite{LMBBR1}.

Thermal Nuclear Reactors may be useful for limited applications, but not for general electric energy generation.  They are useful for marine propulsion \cite{korabli}.  In the author's work on nuclear reactors \cite{GCR}, a case was made that thermal nuclear reactors could be very useful for space propulsion.  The author also made a case against proliferation of thermal nuclear reactors on Earth -- $^{235}$U consumed in these reactors will deplete a fuel resource needed for space transportation \cite[p. 102]{GCR}.  By the time $^{235}$U will be needed for space exploration, most uranium resources may be depleted.

Seawater contains about 4.5 billion tons of uranium in very low concentrations -- 3.3 parts per billion by weight \cite{USea02}.  This resource seems to have the ability to provide enough uranium even for Thermal Nuclear Reactors.  Unfortunately, this uranium can not be extracted at reasonable costs.  Production cost of uranium extracted from seawater is estimated at \$600 to \$1,000 per $kg$ \cite{USea02}.  According to other sources, it would cost \$1,000 to \$1,400 per $kg$ to extract uranium from seawater \cite{USea03}.

\subsection{Thermonuclear Energy}
Thermonuclear energy is generated by the nuclear fusion reaction
\[
^2H+^3H \to ^4He+n+17.59\ MeV.
\]
One gram of fuel generates 270 $GJ$ -- as much as 8.5 tons of coal.
Deuterium $\big($$^2H \big)$ is common in nature -- each ton of ocean water contains 140 $g$ of deuterium \cite[p.11-51]{crc}.  Tritium $\big($$^3H \big)$ is not present in nature.  Tritium is generated from the isotope $^6$Li via the reaction
\[
^6Li+n \to ^4He+^3H.
\]
At first glance, it may seem that one fusion produces only one neutron, which must be consumed in order for one tritium atom to be generated.  In reality, the energetic neutrons produced by nuclear fusion knock out neutrons from other atoms.  Thus, more neutrons are produced than consumed.  On average, one nuclear fusion event, which consumes one tritium atom, produces 1.2 tritium atoms \cite[p.67]{Triton1}.

The stocks of $^6$Li are also virtually unlimited.  Lithium makes up 20 parts per million (ppm) in Earth's Crust by mass \cite[p.14-14]{crc}, and $^6$Li makes up 6.6\% of lithium by mass \cite[p.1-15]{crc}.  Currently, lithium production is 85,000 tons per year, while lithium reserves are about 14 million tons. The price of Li$_2$CO$_3$ was \$17 per $kg$ in 2018.  That is about \$91 per $kg$ lithium \cite[p.99]{minerals2019}.

Nuclear Fusion could have been an almost unlimited source of energy.  Unfortunately, it was a lost opportunity.  According to a 1960 report, there should be about 250 nuclear fusion power plants in Europe in 20 years \cite{NFusion}.
During many decades, many experts believed that nuclear fusion power is 20 years away.
Some people are still optimistic about Nuclear Fusion, while others have given up hope.

One of the main reasons why Nuclear Fusion did not succeed is the lack of funding.
Between the years 1975 and 1982, the average annual budget for fusion power in the USA was \$1 billion per year, after which the funding fell rapidly \cite{NFBudget1}.  Between the years 2000 and 2012, the average annual budget for fusion power in the USA was \$300 million to \$400 million per year \cite{NFBudget2}.  According to a 1976 plan for development of nuclear fusion power, these levels of funding would never achieve a result \cite[p.12]{FPlan}.  In Europe, a giant thermonuclear power station called ITER is being constructed.  It's total cost of \$22 Billion is covered by 35 Nations.  It is supposed to start working in 2035 \cite{NFBudget3}.

In the author's opinion, Nuclear Fusion based Energy Revolution would have succeeded if it had more funding.  Many experts agree \cite{NFFund1,NFFund2,NFFund3,NFFund4}.  Had funding for Fusion Power been at least \$30 Billion per year since 1980, it is likely that the Fusion Power Revolution would have started by the turn of the century.  Between 2010 and 2015, yearly investments into solar energy fluctuated between \$110 billion and \$160 billion.  Investments into wind energy fluctuated between \$80 billion and \$110 billion at the same time \cite{EEnAll}.

\section{Non-Conventional Liquid Fuel Reserves}
\subsection{Heavy Oil and Oil Shale}

Non-conventional oil reserves can not engender an Energy Revolution. First, while these reserves are vast, they are not that vast.  Second, many of these reserves require a huge amount of energy to be extracted.  These reserves can provide a large supply of liquid hydrocarbon fuel and chemical feedstock for applications for which pure electric energy is unsuitable.

Venezuela has about 90 billion tons of recoverable heavy oil is in Orinoco Oil Belt.  This oil is concentrated in sandstone reserves in an area of 50,000 $km^2$.  Heavy oil reserves are at a depth of 150 $m$ to 1.4 $km$.   This oil is much more viscous than conventional oil.  In order to recover this oil, sandstone has to be heated to about 100 $^o$C \cite{VenezuellaOil}.

Alberta, Canada, has about 270 billion tons of bitumen, of which about 180 billion tons should be recoverable.  Bitumen is even more viscous and harder to extract than heavy oil.  Bitumen is present in a mixture with sand, in which sand makes up 90\% to 95\% by weight.
Bitumen can be extracted by two methods.
It can be mined along with the sand, and then processed.
Alternatively, bitumen-sand matrix can be heated in situ to about 120 $^o$C.  Then bitumen can be extracted as crude oil \cite[p.2-10]{ATarS}.

USA and Russia have large reserves of oil shale.  Green River Formation is located in Wyoming, Utah, and Colorado.  It contains about 480 billion tons of recoverable oil in oil shale.  Most oil shale is close to the surface \cite[p.11]{GRiver}.
Bazhenov Formation is in Western Siberia.  It contains 910 billion tons of recoverable oil in oil shale \cite[p.ix-2]{OShale1}. The formation has a thickness of 20 $m$ to 60 $m$ and an overburden of 2.5 $km$ to 3 $km$.  The central part of 600 $km$ by 600 $km$ contains rock with at least 7\% organic carbon \cite[p.22]{Bazh1}.
At least 200 billion tons of recoverable oil in oil shale are contained in places other than Russia and USA \cite{OShale}.

Extracting oil from oil shale is much more energy-intensive than extracting heavy oil or bitumen.  Oil shale does not contain oil -- it contains kerogen.  Kerogen is a complicated mixture of hydrocarbons of organic origin.  When kerogen is heated to 360 $^o$C to 420 $^o$C, it undergoes pyrolysis releasing oil vapour \cite[p.1]{OShaleHC3}.

There are different processes for shale heating in situ. Shell uses vertical wells with inserted electric heaters.  ExxonMobil’s electrofrac$^{\text{TM}}$ process uses fractures filled with electrically conductive material.  This material is kept hot by electric current.  It constantly transfers heat to surrounding rock.    Schlumberger/Raytheon-CF radio-frequency heating uses radio waves to heat shale \cite{OShaleHC3}.

In order to calculate energy requirements for oil extraction, we calculate how much oil can be obtained by heating one ton of rock and how much energy is required to heat one ton of rock.
One ton of Bazhenov Formation kerogen releases about 0.38 tons of oil on heating \cite[p.869]{Bazh2}.  Kerogen makes up  about 11\% of Bazhenov Formation shale by mass \cite[p.ix-2]{OShale1}.  From aforementioned data, it follows that a ton of Bazhenov Formation shale releases about 40 $kg$ of oil on heating.
Green River Formation shale releases 13 $kg$ to 93 $kg$ oil per ton shale on heating.
About 48 billion tons of oil are available from high grade shale, which releases 65 $kg$ to 93 $kg$ oil per ton.
About 110 billion tons of oil are available from medium grade shale, which releases 39 $kg$ to 64 $kg$ oil per ton.
About 320 billion tons of oil are available from poor grade shale, which releases 13 $kg$ to 38 $kg$ oil per ton \cite{GRiver}.

The heat capacity of oil shale is \cite[p.189]{OShaleHC1}
      \[
      C_p=\Big[0.58+0.0038\ T+f_{_O} \big(0.52+0.0023\big)\ T \Big]\ \frac{J}{g\ ^oC}\ ,
      \]
where $f_{_O}$ is the organic matter fraction and $T$ is temperature in $^o$ C.  From the data above, it follows that it takes
      \[
      H=\big[524+381\ f_{_O} \big] \frac{J}{g}
       =\big[146+106\ f_{_O} \big] \frac{kWh\ \text{thermal}}{ton}
      \]
to heat shale from 20 $^o$C to 400 $^o$C.

In Table \ref{T01} below, we present the energy requirements for extraction of oil from shale.
\begin{center}
  \begin{tabular}{|l|l|l|l|l|l|l|l|l|l|}
    \hline
    $kg$ oil per ton shale              &  13   & 20   & 30   & 40  & 60   & 80  & 100  \\
    organic matter fraction $f_{_O}$    &  0.03 & 0.05 & 0.08 & 0.1 & 0.15 & 0.2 & 0.25 \\
    $kWh$ used per $ton$ shale          &  149  & 151  & 154  & 157 & 162  & 167 & 173  \\
    $kWh$ used per $kg$ oil             &  11.5 & 7.6  & 5.1  & 3.9 & 2.7  & 2.1 & 1.7  \\
    \hline
  \end{tabular}
\captionof{table}{Energy Cost of Shale Oil Extraction \label{T01}}
\end{center}
One $kg$ of oil contains 11.6 $kWh$ thermal energy.  Thus, oil extraction from shale is very energy-intensive.

\subsection{Coal Liquefaction}

Liquid hydrocarbon fuel can be obtained from coal.  This technology is energetically expensive, thus it is rarely used in modern world.  Coal liquefaction has a troubled history.
Large scale liquefaction of coal began in Germany during WWII.  ``By the end of the war, in Germany nine indirect and 18 direct liquefaction plant were producing almost 4 million tonnes/year of gasoline, 90\% of German consumption" \cite[p.4]{CoalTL1}.  South Africa had a large coal liquefaction program in the 1980s.  Three plants produced 10 million tons of transportation fuel per year \cite[p.4]{CoalTL1}.

Coal can be transformed into liquid hydrocarbons by two processes.  In \textbf{direct liquefaction} process, hydrogen is reacted with coal to produce liquid fuel.  The reaction takes place at about 300 $atm$ pressure and 470 $^o$C temperature.  The coal feed rate is about 0.5 $ton$ to 0.65 $ton$ per $m^3$ of reactor volume.  Prior to reaction, coal is mixed with about 2\% by mass iron oxide catalyst \cite[p.6]{CoalTL1}.  In \textbf{indirect liquefaction} process, coal is gasified, and the gas is converted to liquid hydrocarbons \cite[p.11]{CoalTL1}.

Indirect liquefaction is the more commonly used process.  Below, we present this process in some detail.
Typical bituminous coal is Illinois \#6 coal.   It contains 11\% moisture, 10\% ash, 3\% sulfur and chlorine, and 76\% organic content.  The empirical formula for organic content is CH$_{_{0.85}}$O$_{_{0.08}}$N$_{_{0.02}}$ \cite[p.19]{FishT5}.
First, coal is heated in the absence of air.  Upon heating, all moisture and all volatile components are released.  Almost all hydrogen, oxygen, nitrogen, and sulfur are released.  Some carbon is also released in volatile components.
Some of released components are condensed as coal tar and water.
Other components remain gaseous at room temperature -- methane, hydrogen, carbon monoxide, carbon dioxide, nitrogen, and hydrogen sulfide.
Coal tar is upgraded to liquid fuel.
Sulfur, water and carbon dioxide are separated.
Methane, hydrogen, and carbon monoxide are used in further gasification and synthesis.
In the less energy-intensive liquefaction, carbon dioxide is released into the atmosphere.
In the more energy-intensive liquefaction, carbon dioxide is also used to produce liquid fuel.

After coal is heated and volatiles escape, the remaining solid is coke, composed mostly of carbon.  This carbon is gasified.  The gasification process consists of passing steam over coke at a temperature of about 1,000 $^o$C, while constantly supplying heat to coke and steam.
The gasification reaction is
      \[
      C+(1+x+y)\ H_2O(g) \ \to \ (1+x)\ H_2+ (1-x)\ CO+ (1+x)\ CO_2+y\ H_2O-\big(131.3-41.2\ x \big)\ \frac{kJ}{mol},
      \]
where $x$ can take any value between 0 and 1.  Usually, $x \approx 0.7$ and $y \approx 2$.

Gas produced by coal contains CO, H$_2$, CO$_2$, and H$_2$O. Some carbon monoxide undergoes the water gas shift reaction:
    \[
    CO+H_2O \to CO_2+H_2
    \]
The above reaction can also go backward.
After the gas is cooled, carbon dioxide and water are separated.  Water is condensed, while carbon dioxide is removed by a chemical process \cite{FishT5,FishT3,FishT2}.

The liquefaction reaction is
      \[
      CO+ (2+x)\ H_2 \to CH_{(2+2x)}+H_2O,
      \]
where $x$ is usually between $0.2$ and $0.5$.  This reaction is called \textbf{Fischer-Tropsch process}.  The liquefaction reaction takes place at temperature 260 $^o$C to 350 $^o$C and pressure of 50 $atm$ to 70 $atm$ \cite[p.12]{CoalTL1}.
Fischer-Tropsch process can use several types of catalysts.  Ruthenium is the best catalyst \cite{Ruthenium}.

Average cost of operating Fischer-Tropsch process plants is \$144,000 per barrel per day, operating capacity \cite[p.9]{GasTL1}.  This is \$9.4 per gallon per year operating capacity.  In planned plants, the costs are up to 2.5 times lower \cite[p.12]{GasTL1}.  For a typical synthetic fuel plant, the cost breakdown is the following: gasification 65-70\%, Fischer-Tropsch synthesis 21-24\%,  upgrading to fuels 9-19\% \cite[p.6]{FT1}.  Usual costs for gas-to-liquid plants are \$60,000 to \$80,000 per barrel per day, while best costs are \$45,000 per barrel per day \cite[p.10]{FT1}.

In modern coal liquefaction, heat and electrical energy for the process is supplied by coal.  In this process, one ton of coal yields an average of 0.1-0.26 tons of liquid fuel \cite{FishT4}. In a Fischer-Tropsch process plant, where all of the thermal energy and motive power is provided externally, one ton of coal should produce $490\ kg$ of fuel oil.

The lower heating value of the fuel obtained by Fischer-Tropsch process is
$H_{_\text{fuel}}= 44.2\ kJ/g$.
The lower heating value of typical coal is  $H_{_\text{coal}}=25.9\ MJ/kg$ \cite[p.19-20]{FishT5}.
The \textbf{lower heating value efficiency} of the process is
      \[
      \frac{490\ kg}{1000\ kg}\ \cdot \
      \frac{H_{_\text{fuel}}}{H_{_\text{coal}}}\ 100\%=84\%.
      \]
Lower heating value efficiency is the fraction of fuel in feedstock which is retained in the product.  The lower heating value efficiency of the state of the art coal liquefaction plants is 33\% to 55\% \cite[p.871]{FishT3}.   Most projected coal liquefaction plants have lower heating value efficiencies of 47\% to 52\% \cite[p.30]{FishT2}.
Given that state of the art plants obtain heat and electric energy from coal itself, this efficiency is expected.

\subsection{Energy-Intensive Coal Liquefaction}

In a more energy-intensive coal liquefaction, excess hydrogen is obtained by water electrolysis.  All carbon present in coal is converted to liquid propellant.  The cumulative liquefaction reaction for coal organic content is
    \[
    CH_{_{0.85}}O_{_{0.08}}N_{_{0.02}}+ 0.81\ H_2 \to CH_{_{2.3}}+0.08\ H_2O+0.01\ N_2,
    \]
where $CH_{_{2.3}}$ is the hydrocarbon fuel.  In this subsection, we estimate electric energy requirements for obtaining one kilogram of hydrocarbon fuel by this process.  There are three primary energy expenditures in energy-intensive coal liquefaction.  First, electric energy is used in hydrogen generation via water electrolysis.  Second, electric energy is used in coal gasification.  Third, electric energy is used in several axillary tasks such as compressing coal gas to 70 $atm$ pressure, separating CO$_2$ from gas feed and reheating it with hydrogen, and various other tasks.

The mass of excess hydrogen needed to obtain 1 $kg$ of hydrocarbon fuel is
    \[
    \frac{\text{Molar mass }\big(0.81\ H_2\big)}
    {\text{Molar mass }\big(CH_{_{2.3}} \big)}\ 1\ kg=
    \frac{2 \cdot 0.81}{12+2.3}\ 1\ kg=0.113\ kg.
    \]
Electric energy consumption for hydrogen from electrolysis of water is about 55 $kWh/kg$ -- considerably greater than the theoretical minimum of 33.6 $kWh/kg$ \cite{Electrolysis,Electrolysis1}.  Electric energy needed to produce electrolytic hydrogen needed for production of 1 $kg$ of hydrocarbon fuel is
    \[
    0.113\ kg\ \frac{55\ kWh}{1\ kg}=6.2\ kWh.
    \]

We use Rocket Propulsion Analysis (RPA) software \cite{RPA} to calculate chemical equilibrium during coal gasification.  Even though RPA software is best suited for calculating chemical equilibrium within rocket engines, it is also suitable for calculating chemical equilibrium within gasification process.  In order to produce 1 $kg$ of hydrocarbon fuel, we must gasify
    \[
    \frac{\text{Molar mass }\big(CH_{_{0.85}}O_{_{0.08}}N_{_{0.02}}\big)}
    {\text{Molar mass }\big(CH_{_{2.3}} \big)}\ 1\ kg=
    \frac{12+0.85+16 \cdot 0.08+14 \cdot 0.02}{12+2.3}\ 1\ kg=1.01\ kg
    \]
of coal organic content.
According to RPA calculation we have performed, 1.01 $kg$ of coal organic content can be gasified by 4.04 $kg$ water vapor at 2,050 $^o$C.  In order to heat 4.04 $kg$ water vapor to this temperature, we need 5.4 $kWh$ electric heat.  The evaporation of water uses waste heat.

Calculating electric energy requirement for axillary tasks is beyond the scope of this work.  Nevertheless, assuming 5.0 $kWh$ electricity per $kg$ is a generous allowance for this electricity.  Adding all components, we obtain 17 $kWh$ per $kg$ synthetic fuel produced.

The overall material cost of 1 $kg$ synthetic fuel would be about 17 $kWh$ electric energy and about 1 $kg$ coal.  Such process would be profitable only in areas, where vast amounts of inexpensive energy are available.  Energy-intensive coal liquefaction plants may be set up in the Sahara Desert.  These plants would use solar power and coal from Europe to produce liquid fuel.

Direct coal liquefaction would only use considerably less energy.  It would need the energy only to generate hydrogen by electrolysis and to compress it to 500 $atm$ pressure.  The overall material cost of 1 $kg$ synthetic fuel would be about 7 $kWh$ electric energy and about 1 $kg$ coal.  The problem with direct coal liquefaction is high capital cost \cite{DCL1}.

\subsection{Additional Uses of Energy}
\subsubsection{Hydrocarbon Fuel From Air}
Hydrocarbon fuel can literally be produced out of air and water.  First, carbon dioxide is captured from the air.  Second, hydrogen is produced by water electrolysis.  Third, carbon dioxide is reduced to carbon monoxide by hydrogen.  Fourth, CO and H$_2$ are reacted to produce hydrocarbon fuel.  The process is the most energy intensive.  Producing 1 $kg$ hydrocarbon fuel would cost 33 $kWh$ \cite{CO2-1}.
According to some detailed studies, more advanced technology could produce hydrocarbon fuel out of air and water at a cost of 21 $kWh/kg$ \cite[p.75]{CO2-4}.  Nevertheless, such technologies may be very expensive.

Technology to produce hydrocarbon fuel from air, water, and electric energy exists now.  This technology is not used for two reasons.  First, it is cost prohibitive.  Second, there are other sources of hydrocarbon fuel available at present.  Solar Power Revolution and rapid increase in demand of all types of energy is likely to change both aforementioned conditions.

\subsubsection{Animal Feed Production}
As World population and material standard of living grows, global requirements for meat and fish production will also grow.  Increased meat production will facilitate increased demand for animal feed.

Protein rich feed can be produced from hydrogen and methane by microorganisms.  In one project, production of 1 $kg$ of feed would require 0.55 $kg$ of hydrogen and 0.14 $kg$ of ammonia in addition to oxygen and carbon dioxide \cite[p.6]{Protein1}.  The total amount of hydrogen needed both for feedstock hydrogen and ammonia production is 0.57 $kg$.  Recalling that each $kg$ of hydrogen requires 55 $kWh$ electric, we obtain the energy requirements for production of 1 $kg$ feed:
    \[
    0.57\ kg \cdot 55\ kWh/kg=31\ kWh.
    \]
This feed contains 70\% protein \cite[p.7]{Protein1}.
Other reactors use 0.44 $kg$ hydrogen and 0.13 $kg$ ammonia to produce 1 $kg$ of feed, 71\% of which is protein \cite{Protein3,EtoFood2}.  In this reactor, 0.46 $kg$ hydrogen or 25 $kWh$ is used to produce 1 $kg$ of feed.  Methane-digesting reactors use 1.25 $kg$ methane to produce 1 $kg$ feed \cite{Methane3}.  Some protein produced may be used as food for human consumption \cite{EtoFood2,EtoFood1}.

Other foods, such as sugars, can also be artificially produced from water, carbon dioxide, and electric energy.  Under ideal conditions, production of this food would cost about 10 $kWh$ per $kg$ \cite{Sugar1}.  Real costs should be considerably higher, thus we will use 25 $kWh/kg$ as tentative cost for all animal feed.

One $kg$ protein or carbohydrate contains 4.2 $Mcal$ or 4.9 $kWh$ of food energy \cite[p.40]{ProtCal}.  The energy efficiency of producing basic food or feed out of water, carbon dioxide, and electric energy is
    \[
    \frac{\text{Energy content of 1 $kg$ food}}
         {\text{Energy used to produce 1 $kg$ food}}\ \cdot\ 100\% =
    \frac{4.9\ kWh}{31\ kWh}\ \cdot\ 100\% = 16\%.
    \]

Production of meat and aquaculture requires large amount of feed.
Producing
1 $kg$ beef requires 25 $kg$ feed,
1 $kg$ lamb requires 15 $kg$ feed,
1 $kg$ pork requires 6.4 $kg$ feed, and
1 $kg$ poultry requires 3.3 $kg$ feed.
One $kg$ of feed protein can produce
0.25 $kg$ egg protein, or
0.24 $kg$ milk or butter or cheese protein, or
0.20 $kg$ poultry protein, or
0.085 $kg$ pork protein, or
0.063 $kg$ lamb protein, or
0.038 $kg$ beef protein.
One feed calorie can produce
0.24 milk calories, or
0.19 egg calories, or
0.13 poultry calories, or
0.086 pork calories, or
0.044 lamb calories, or
0.019 beef calories \cite{Myaso}.

We have technology which can produce meat ``out of thin air".  A factory would convert carbon dioxide, water, and electricity into animal feed.  A farm requiring no pasture would raise livestock on artificial feed.  Using the feed requirements mentioned in the previous paragraph and 25 $kWh/kg$ as energy cost of feed, we obtain the energy cost of meat.
Producing
1 $kg$ beef requires 630 $kWh$,
1 $kg$ lamb requires 380 $kWh$,
1 $kg$ pork requires 160 $kWh$, and
1 $kg$ poultry requires 83 $kWh$.

One advantage of production of meat using artificial feed is lower land requirements compared to conventional agriculture.
Currently, it takes at least 24 $m^2$ of crop land and pasture to produce 1 $kg$ beef per year \cite[p.7]{Proteinn}.  According to other estimates, the minimal use of crop and pasture land is
15 $m^2$ to produce 1 $kg$ beef per year,
8 $m^2$ to produce 1 $kg$ pork per year, and
5 $m^2$ to produce 1 $kg$ poultry per year \cite[p.1208]{Proteinn}.
Below we calculate the land requirements for meat production using artificial feed.  Most deserts receive annual solar irradiation of at least 2,000 $kWh/m^2$ \cite[p.55]{FSolar}.  A 24\% efficient solar power station would convert this energy into 480 $kWh/m^2$ electric.  Given energy requirements for meat production presented above, we calculate land requirements for producing different kinds of meat from artificial feed.
These requirements are
1.3 $m^2$ to produce 1 $kg$ beef per year,
0.8 $m^2$ to produce 1 $kg$ pork per year, and
0.17 $m^2$ to produce 1 $kg$ poultry per year.
Land requirements for producing meat via artificial feed are about 11 to 29 times lower than land requirements for most efficient state of the art farms.

Currently, technology for producing animal feed from water, carbon dioxide, and electrical energy is not used.  The reasons are similar to those for which hydrocarbon fuel is not manufactured from the same ingredients.  First, it is cost-prohibitive.  Second, there is still enough conventional feed to raise livestock and poultry.  By the end stages of the Solar Power Revolution, energy will become plentiful.  Global requirements for food and meat may become unsustainable using conventional feed.

Global meat production grew from about 70 million tons in 1961 to about 320 million tons in 2014.  Between 1961 and 2014, global annual meat consumption per capita grew from 20 $kg$ to 43 $kg$.  An average American consumes 110 $kg$ of meat per year, including about 37 $kg$ beef \cite{Myaso}.
By the time global per capita meat consumption reaches USA 2020 levels, the World population may be 11 billion people \cite{pop2050}.
Global annual meat consumption would be 1.2 billion tons, including 410 million tons of beef.
Sustaining a meat industry on this scale would require consumption of about 16 billion tons of dry animal feed per year.  Production of this feed would require an  expenditure of 400 $PWh$ electric per year.  As we have discussed in Subsection 3.1, such energy requirements should be more than feasible at the later stages of the Solar Power Revolution.

Global fish catch grew from about 30 million tons in 1950 to over 125 million tons in 1997.  Since then, global fish catch has slightly declined to about 110 million tons \cite{Fish1}.
Global fish catch can neither grow, nor remain at current levels.  In order to save fauna, fishing must decline and be replaced by aquaculture.  About 33\% of global fish stocks are overfished, about 60\% are fished to a maximum extent, and only 7\% are under fished \cite[p.40]{Aquaculture}.

Aquaculture production also requires a lot of feed energy.  One $kg$ of feed protein can produce an average of 0.2 $kg$ fish protein, and one feed calorie can produce an average of 0.1 fish calories \cite[p.5]{Pig1}.
Between 1995 and 2016, World aquaculture production grew from 24 million tons to 80 million tons \cite[p.27]{Aquaculture}.
Global fish consumption is close to North American fish consumption -- 20 $kg$ per capita per year \cite[p.72]{Aquaculture}.  Aquaculture is projected to grow moderately in the coming decades \cite[p.184]{Aquaculture}.

\section{The Use of New Energy}
\subsection{Raising Global Material Standards to US Material Standards of 2020}
First, solar and wind electric energy would displace fossil fuels in many areas.
As solar power and wind power would become the main sources of electricity, fossil-fuel based power stations would disappear.  Electric heat pumps would displace fossil fuels for space heating, \cite{HPump}.  Electricity would provide heat for industrial processes.  Liquid hydrocarbon fuels would still be used for transportation and chemical feedstock.

Second, solar and wind power would enable the growth of Global material living standard to US material living standard.  That would involve the growth of Global energy consumption per capita to US energy consumption per capita.  Below, we calculate energy requirements for this transition.

As we have mentioned in Section 2.3, in 2018 USA has used about 24 $PWh$ thermal in fossil fuels, and generated 4.0 $PWh$ electric energy.  As electricity would replace fossil fuels for space heating and industrial heat, electricity consumption would rise to about 6 $PWh$.
Additionally, in USA 2018, 8.4 $PWh$ thermal worth of liquid hydrocarbons was used in transportation and 1.7 $PWh$ thermal worth of hydrocarbons was used as chemical feedstock.  The mass of hydrocarbons annually used for transportation and as chemical feedstock in the USA is about 870 million tons.

In early 2020, US population was 331 million, while World population was 7,760 million \cite{Worldometer}.  According to United Nations, World population should reach 11 billion by the second half of 21$^{\text{st}}$ century \cite{pop2050}.  In order for global population to live at material living standards of USA 2020, the World would have to consume
    \[
    6\ PWh\ \cdot \frac{\text{World population second half of 21$^{\text{st}}$ century}}
    {\text{US population 2020}}=200\ PWh
    \]
electricity, which is 7.8 times the modern consumption (Section 2.4).

The World would have to consume
    \[
    0.87\ GT\ \cdot \frac{\text{World population second half of 21$^{\text{st}}$ century}}
    {\text{US population 2020}}=29\ GT
    \]
of liquid hydrocarbon fuel, where $GT$ denotes a billion tons.  In 2017, the World consumed 4.5 billion tons of oil and 2.8 billion tons of natural gas \cite[p.6]{WEnrg18}.  In order to expand liquid hydrocarbon production to 29 billion tons per year, non-conventional fuel reserves would have to be used widely.  First, heavy oil and shale oil resources would have to be tapped in.  Then, coal liquefaction would have to be used.  Finally, hydrocarbon fuel would have to be generated by CO$_2$ capture.  These processes would have to consume up to
    \[
    2.9 \cdot 10^{13}\ kg \text{ hydrocarbon} \cdot
    \frac{\text{21 kWh \text{ electric} \cite[p.75]{CO2-4}}}{\text{kg \text{ hydrocarbon}}}=610\ PWh \text{ electric}
    \]
per year.

As we have mentioned in Subsection 4.4, global meat industry is likely to consume another 400 $PWh$ electric per year.  Along with 200 $PWh$ a year in electricity, 610 $PWh$ a year in hydrocarbon fuel production, and over 100 $PWh$ a year in unforeseen energy expenses, the World in which 11 billion people live at high material standard of living would require 1,300 $PWh$ per year.  This power should be available by the end of the century.  As we have mentioned in Subsection 3.1, this is still only 8\% of Solar Energy potential.

\subsection{Beyond US Material Standards of 2020}
Sometime at the end of the 21$^{\text{st}}$ century, Global electricity, motor fuel and meat consumption per capita should be the same as the corresponding parameters in USA 2020.  As we have calculated in Subsection 5.1, World electricity generation at that point would be 1,300 $PWh$ per year.  Given a population of 11 billion, electric energy generation per capita would be about 120,000 $kWh$ per year.  That is more than in modern USA, since neither hydrocarbon fuel nor animal feed is generated from carbon dioxide, water, and electric energy in 2020.

Average income per capita and material standard of living in the World at that time should be much higher than the corresponding standards in USA 2020.  First, energy efficiency increases as technology matures \cite{EE01}.  Second, there is constant improvement in electronics, medicine, and consumer products.  Third, robotic technology is rapidly advancing \cite{Robots}.  Additive manufacturing holds great promise \cite{AddM1,AddM2,AddM3}.

Nevertheless, after that point, Solar Energy Revolution would continue until World energy consumption reaches the full potential of 23,000 $PWh$ per year discussed in Section 3.1.  That would enable Earth to support a greater population -- perhaps 50 billion.  Electricity production per capita would be 460,000 $kWh$ per year.  Abundance of energy along with very advanced manufacturing and robotic technology would enable all of Earth's population to have high material standard of living.

Another likely effect of Energy Revolution would be further connection of the Globe.  By 2020, the Internet has connected the Globe in Information Space.  Now it is possible to talk to people in any place on Earth.  Many modern people have anonymous friends all over the Globe.
Nevertheless, long distance travel is still expensive.  Abundance of energy may enable Humankind to connect the World with a system of Global Rapid Transit.  One type of transit being considered consists of Magnetic Levitation (MagLev) trains \cite{GlobalTram1}.  A 25,000 mile MagLev network has been proposed for the USA \cite{MagLev01}.  MagLev trains designed in 2002 would have capability of travelling at 450 $km/h$ \cite{MagLev05}.

During late 19$^{\text{th}}$ and early 20$^{\text{th}}$ centuries, most American and some European cities grew with areas called streetcar suburbs.  These suburbs had benefits of both the city center and of suburban areas.  On one hand, the city center could be easily reached from these suburbs by electric streetcars.  On the other hand, streetcar suburbs were much more scarcely populated than city centers.  They had much lower real estate prices \cite{SSuburb1,SSuburb2,SSuburb3}.  A system of MagLev train lines can turn Earth into a Global City.  The envisioned Global City is called Ecumenopolis -- which is ``a city made of the whole world" in Greek \cite{Ecumenopolis1,Ecumenopolis2}.

\subsection{Kardashev 2 Civilization -- the Final Frontier}
In Subsection 5.2, we have described some material aspects of a society at the end of the Solar Power Revolution.  It is a global Ecumenopolis connected not only by Internet but also by rapid transit.  This Ecumenopolis contains about 50 billion people.  All people have material standard of living much higher than an average American in 2020.  Electrical energy production is about 23,000 $PWh$ per year -- close to an absolute maximum possible without destroying Earth's ecosystem.

A civilisation which uses Solar Energy striking Earth to a maximal possible extent is classified as Kardashev Type 1 Civilization \cite{kard}.  This level of achievement is remarkable by 2020 standards.  Nevertheless, the Solar System contains resources to support a much more extensive civilization.

Kardashev Type 2 Civilization would consist of colonized Solar System.  This civilization will vastly exceed any civilization which has existed on Earth -- even Kardashev Type 1 Civilization.  Material resources available on planets, satellites, and asteroids are vastly greater than those available within Earth's crust.  Solar power available in Space exceeds solar power available on Earth by a factor of a billion.  The Solar System has resources to provide very high material standard of living for 10 quadrillion people \cite{quadrillion}.  All of these people will have material standard of living much higher than people living in Kardashev Type 1 Civilization.

In Kardashev Type 2 Civilization, humans will live on billions of large habitats orbiting the Sun.  Each of these habitats will harvest solar energy and produce all necessary food, drinking water, and oxygen needed for humans and animals.  This concept is called the Dyson Sphere \cite{Dyson}.  It was first envisioned by Konstantin Tsiolkovsky back in 1903 \cite{Tsialkovski}.  During the 1970s, many elaborate models of Kardashev Type 2 Civilization were published \cite{habitats,1974}.

When and how will Solar System Colonization begin?  At this point, we can not know.  On one hand, Elon Musk believes that colonization of Solar System can start in the 2020s \cite{Musk1,Musk2,Musk3} -- before the beginning of Solar Power Revolution.  On the other hand, Solar System Colonization may begin during or after the Solar Power Revolution.  The only thing we can say is that every step in technological development makes the next step easier and more likely.

\section{Conclusion}

In this work, we have described a likely scenario of Solar Power Revolution, and corresponding changes in material living standards of Global Civilization.  Energy use by Humankind has greatly expanded since the beginning of the Industrial Revolution.  As mentioned in Subsection 2.4, Global consumption of energy in 2017 was 147 $PWh$ thermal.  That year, 25.6 $PWh$ electric was generated.   As mentioned in Subsection 3.1, at the end of Solar Power Revolution, Global consumption of energy will be 23,000 $PWh$ electric per year.  Wind may be an important source of energy at early stages of Solar Power Revolution, but overall potential of wind is much lower than that of solar.  Nuclear and thermonuclear energy held great promise, but this promise did not materialize.

Solar Power Revolution and rapid growth of electricity generation is likely to bring profound changes in energy use as well as general material standard of living.
First, fossil fuel will be phased out from electricity generation.
Second, many functions of fossil fuel -- such as space heating and industrial process heating, will be performed by electricity.  Nevertheless, hydrocarbon fuel will still be needed for motor vehicles and as feedstock in chemical industry.
Third, electric energy will be used to produce hydrocarbon fuel and chemicals.  Electric energy will be used to extract heavy oil and produce oil from oil shale.  Then, electric energy will be used to produce liquid hydrocarbons from coal.  Finally, electric energy will be used to produce both liquid hydrocarbons and animal feed from carbon dioxide and water.

At the end of Solar Power Revolution, we will have Kardashev Type 1 Civilization.  This civilization will have a population of about 50 billion, with a material standard of living much higher than an average American in 2020.  Electric energy production will be 23,000 $PWh$ per year, which is 460,000 $kWh$ per year per capita.  The World will be connected by a network of rapid transit.

The final step for Humankind will be colonization of Solar System and building of Kardashev Type 2 Civilization.  This civilization will use the vast resources of energy and material available within the Solar System.  Kardashev Type 2 Civilization can sustain about 10 quadrillion people at material standards much higher than those of Kardashev Type 1 Civilization.  Most people will live on billions of spaceships.  Earth will remain a cultural center with population of a few hundred billion.

We know the ultimate resources of Solar Power which can be harvested on Earth without destroying the global ecosystem -- thus we can make relatively accurate predictions about material resources of Global Civilization at the end of the Solar Power Revolution.
We can also estimate the material and energy resources contained within the Solar System.  Thus, we can estimate population and material living standard of Kardashev Type 2 Civilization.

We do not know, when Solar Power Revolution will begin and how long it will take.  In the author's opinion, it should begin around 2040 and be completed by around 2120.  We do not know, and perhaps we can not know what kind of unexpected problems may face Global Society during the Solar Power Revolution.  The same is true of discoveries and technologies which may accelerate the Solar Power Revolution.  Most speculations on either subject are unscientific.  The Future remains mostly a mystery.

\end{document}